\documentclass[aps,prl,twocolumn,preprintnumbers,groupedaddress,superscriptaddress,floatfix,tightenlines,reprint]{revtex4-1}
\usepackage{mathrsfs,natbib}
\usepackage{graphics,epsfig,color, graphicx}
\usepackage{verbatim}

\usepackage{graphicx,epsf,amssymb,bbm,amsbsy,amsfonts,amssymb,amsmath}
\usepackage{hyperref}

\newcommand{\eq}{\begin{equation}}
\newcommand{\eqe}{\end{equation}}

\newcommand{\Tr}{\mbox{Tr}\,}

\newcommand{\eqa}{\begin{eqnarray}}
\newcommand{\eqae}{\end{eqnarray}}

\hypersetup{
    colorlinks=true,       
    linkcolor=red,          
    citecolor=blue,        
    filecolor=magenta,      
    urlcolor=blue           
}

\begin{document}

\title{Casimir energy of confining large $N$ gauge theories}

\preprint{PUPT-2469, IPMU14-0254}
\preprint{FTPI-MINN-14/22, UMN-TH-3347/14}

\author{G\"ok\c ce Ba\c sar}
\email{basar@tonic.physics.sunysb.edu}
\affiliation{Department of Physics and Astronomy,
Stony Brook University,
Stony Brook, New York 11794, USA}

\author{Aleksey Cherman}
\email{acherman@umn.edu}
\affiliation{Fine Theoretical Physics Institute, Department of Physics, University of Minnesota, Minnesota, MN 55455, USA}

\author{David A.\ McGady}
\email{dmcgady@princeton.edu}
\affiliation{Department of Physics, Jadwin Hall Princeton University Princeton, NJ 08544, USA.}

\author{Masahito Yamazaki}
\email{masahito.yamazaki@ipmu.jp}
\affiliation{Institute for Advanced Study, School of Natural Sciences, Princeton NJ 08540, USA}
\affiliation{Kavli IPMU (WPI), University of Tokyo, Kashiwa, Chiba 277-8586, Japan}

\begin{abstract}
Four-dimensional asymptotically-free large $N$ gauge theories compactified on $S^3_R \times \mathbb{R}$ have a weakly-coupled confining regime when $R$ is small compared to the strong scale.   We compute the vacuum energy of a variety of confining large $N$ non-supersymmetric gauge theories in this calculable regime, where the vacuum energy can be thought of as the $S^3$ Casimir energy.  The $N=\infty$ renormalized vacuum energy turns out to vanish in all of the large $N$ gauge theories we have examined, confirming a striking prediction of temperature-reflection symmetry.   
\end{abstract}


\maketitle

{\bf Introduction}---In typical quantum field theories (QFTs) with a mass gap $M_0 > 0$, the mass $M$ of the heaviest particle species sets the natural size of the vacuum energy $V \sim M$.   The Standard Model (SM) contains a variety of gapped sectors, and the electron contribution to the vacuum energy density $\mathcal{O}(m_{e}^4) \sim 6 \times 10^{-2}\,\mathrm{MeV}^4$ is already much larger than the value $\sim 1 \times 10^{-36}\,\textrm{MeV}^4$ inferred from the accelerating expansion of the universe\cite{Frieman:2008sn}.  The apparent need to fine-tune $V$ against $M$ is the cosmological constant problem.  

In gapped QFTs the only known mechanism which naturally gives $V=0$ is linearly realized supersymmetry (SUSY).  But if the SM  is the low energy limit of a SUSY QFT, SUSY must be broken at some scale $\mu_{\rm SUSY} \gg m_e$ (see e.g.\ \cite{Khachatryan:2011tk,*Aad:2011hh}), and the cosmological constant problem remains severe.  This strongly motivates a search for other mechanisms that would force $V$ to vanish. 

If a QFT has a finite number of particle species it seems difficult to escape the conclusion that $V \sim M$, but what sets the scale of $V$ if there are an infinite number of species with increasing masses\footnote{See \cite{Moore:1987ue,*Dienes:1994np,*Dienes:1995pm,*Dienes:2001se,*Kachru:1998hd} for earlier discussions of this type of question.}?  This is the situation in weakly-coupled string theories and in confining large $N$ gauge theories, which are believed to have a dual string description\cite{tHooft:1973jz}.    In this paper we compute the vacuum energy of a variety of \emph{non-supersymmetric} $SU(N)$ gauge theories at $N=\infty$, including pure Yang-Mills theory.   The calculations are done using a compactification of spacetime to $S^3_R \times S^1_{\beta}$, where these theories develop an analytically tractable confining regime\cite{Aharony:2003sx} if the $S^3$ radius $R$ is much smaller than the strong scale $1/\Lambda$, and if the temperature $T = 1/\beta$ is below a critical value.  
In this regime $V$ is simply the Casimir energy $C$ of the theory on $S^3 \times \mathbb{R}$.    It was recently observed\cite{Basar:2014mha}  that temperature-reflection ($T$-reflection) symmetry predicts that the vacuum energy associated with the $N=\infty$ spectrum of these confining theories should \emph{vanish}.  

Our calculations confirm this prediction. Since the result holds in a variety of large $N$ gauge theories, it seems unlikely to be an accident.  It is possible that confining gauge theories have emergent symmetries in the large $N$ limit which force $V$ to vanish.

{\bf $T$-reflection}---For  QFTs on $S^3_R \times S^1_{\beta}$ the spectrum of single-particle excitations is discrete, and in our cases of interest the partition function can be written as  
\begin{align}
-\log Z(\beta) &= -V_0 \, \beta  \mathcal{V} +\sum_{\pm,n=1}^{\infty}\left[ \pm \frac{\beta}{2} d^{\pm}_{n} \omega^{\pm}_{n} \right]\nonumber\\
 &+ \sum_{\pm, n=1}^{\infty}\left[ \pm d^{\pm}_{n} \log\left(1 \mp e^{-\beta \omega^{\pm}_{n}}\right) \right]
 \label{eq:FiniteVolumeZ}
\end{align}
where $V_0$ is the bare vacuum energy, $\mathcal{V}$ is the spatial volume, and  $\omega^{\pm}_n, d^{\pm}_n$ are the energies and degeneracies of bosonic (+) and fermionic (-) states.  We study theories where $\omega^{\pm}_n$ only depends on the scale $R$.  The sum in the upper line is UV divergent and must be regulated and renormalized to obtain a physical expression.  The renormalized contribution explicitly depends on $R$ and is the Casimir energy. In \cite{Basar:2014mha} we noted that one can also formally define the quantity $Z(-\beta)$ by sending $\beta \to -\beta$ in  \eqref{eq:FiniteVolumeZ}:
\begin{align}
 \label{eq:FiniteVolumeZminus}
 &-\log Z(-\beta)= V_0 \, \beta  \mathcal{V} +\log(-1)\sum_{n=1}^{\infty} d^{+}_{n}  \\
&+\sum_{\pm,n=1}^{\infty}\left[ \pm \frac{\beta}{2} d^{\pm}_{n} \omega^{\pm}_{n}\right] 
 + \sum_{\pm, n=1}^{\infty}\left[ \pm d^{\pm}_{n} \log\left(1 \mp e^{-\beta \omega^{\pm}_{n}}\right) \right]\nonumber
\end{align}
Of course, $Z(-\beta)$ also has UV divergences, and requires the same type of regularization and renormalization as $Z(\beta)$.  With renormalized expressions for both $Z(\beta)$ and $Z(-\beta)$ in hand it can be shown that there is a $T$-reflection symmetry\cite{Basar:2014mha}
\begin{align}
Z(\beta) = e^{i\gamma} Z(-\beta)
\label{eq:Treflection}
\end{align}
where $\gamma = \pi\, \mathrm{Finite}[\sum_{n=1} d^{+}_n]$~\footnote{There are branch cuts ambiguities in the definition of $\gamma$, and one can choose it to live in $[0,\pi]$. }, provided that the $R$-independent part of the vacuum energy from $V_0$ is set to zero. Hence \eqref{eq:Treflection} holds only if the renormalized vacuum energy $V$ coincides with the Casimir energy $C=1/2\sum_{\pm,n} d^{\pm}_{n}\omega^{\pm}_{n}$. 
For instance, on $S^3_{\, R} \times S^1_{\beta}$, Eq.~\eqref{eq:Treflection} holds for a real conformally-coupled scalar field when $V = 1/(240R)$ and $\gamma=0$, while for an Abelian vector field $T$-reflection holds with $V = 11/(120R)$ and $\gamma=\pi$.

\begin{figure*}[h!t]
  \centering
\includegraphics[width=0.9\textwidth]{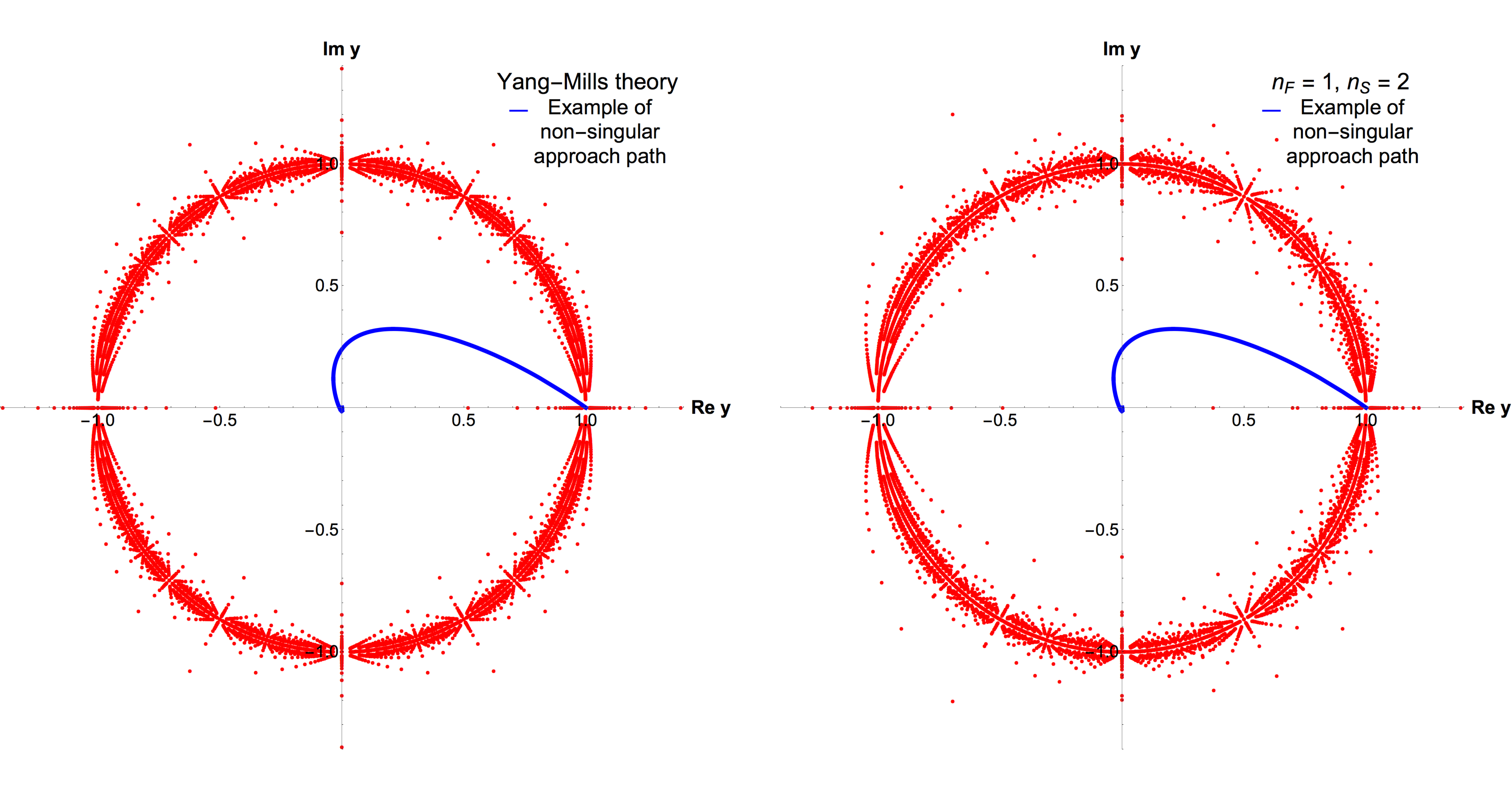}
  \caption{(Color Online.) Structure of singularities (red dots) coming from the first $45$ terms in \eqref{eq:ZST} in the large $N$ confining-phase partition functions of gauge theories with adjoint matter on $S^3 \times S^1$, in the complex plane for $y = e^{-\beta/(2R)}$.  The blue curve is an example of a path from $y=0$ to $y=1$ which does not pass through any singularities.  {\bf Left:}  Yang Mills ($n_F=0, n_S=0$) theory.  {\bf Right:}  Gauge theory with $n_F=1, n_S=2$. }
  \label{fig:YMPlots}
\end{figure*}

{\bf Non-abelian gauge theories on $S^3_R \times S^1_{\beta}$}---We analyze $SU(N)$ gauge theories with $n_F$ adjoint Majorana fermions and $n_S$ real adjoint scalars on $S^3_R \times S^1_{\beta}$.  For moderate $n_F, n_S$, these theories are asymptotically free with a strong scale $\Lambda$, and are weakly coupled if $R\Lambda \ll 1$.   Indeed, in the $\Lambda R \to 0$ limit where we will work, the 't Hooft coupling $\lambda$ goes to $0$, and these theories develop a conformal symmetry at the microscopic level.  However, no matter how small $\lambda$ becomes, the Gauss law constraint on the compact manifold $S^3$ only allows color-singlet operators to be part of the space of finite-energy states, and these operators must include one or more color traces.  

As explained in detail in \cite{Aharony:2003sx} (see also \cite{Sundborg:1999ue,Polyakov:2001af}) in the large $N$ limit such theories have at least two distinct phases.  In particular, there is a low temperature confining phase, dominated by the dynamics of an infinite number of stable single-trace hadronic states, and a mass gap of order $1/R$.  The confined phase has a free energy scaling as $N^0$ and unbroken center symmetry.   

In this paper, we focus on the weakly-coupled large $N$ confining phase, since we wish to compute the vacuum energy of the theory on $S^3 \times \mathbb{R}$. The Casimir energy is dictated by the energies and degeneracies of the states of the theory, which are in turn encoded within the thermodynamic partition function, $Z(\beta) = \Tr\, e^{-\beta H}$. We shall use the spectrum of states in the $N=\infty$ limit to compute the Casimir energy.  Before proceeding to the vacuum energy computation, we review and expand on the remarks in \cite{Basar:2014mha} concerning the $T$-reflection properties of $Z(\beta)$ in $N=\infty$ confining gauge theories on $S^3 \times S^1$.

In large $N$ confining phases, the physical excitations are created by single-trace operators which generate the physical single-particle states.  Hence the thermodynamic partition function associated to the spectrum of excitations on $S^3 \times \mathbb{R}$ is given by \eqref{eq:FiniteVolumeZ} with the spectral data $\omega^{\pm}_n, d^{\pm}_n$ taken from the single-trace thermodynamic partition function\cite{Aharony:2003sx}
\begin{align}
\label{eq:ZST}
-Z_{\rm ST}(\beta)  = \sum^{\infty}_{k=1} &\frac{\varphi(k)}{k} \log\left[1-z_V(x^{k}) - n_S z_S(x^{k})\right.\\
&\left. +(-1)^k n_F z_F(x^{k}) \right] =: \sum_{n=1}^{\infty} d_n y^n \nonumber
\end{align}
where $\varphi(k)$ is the Euler totient function, $x = e^{-\beta/R}$, $y=x^{1/2}$, states with even/odd labels $n$ are bosons/fermions, and
\begin{align}
z_S(x) &= \frac{x^2+x}{(1 -x)^3},\,\,z_F(x) = \frac{4x^{3/2}}{(1 -x)^3},\,\, z_V(x) = \frac{6 x^2-2 x^3}{(1-x)^3} \nonumber
\end{align}
are the so-called single-letter partition functions for respectively the conformally-coupled real scalar, Majorana fermion and Maxwell vector fields on $S^3$.

To relate this to \eqref{eq:FiniteVolumeZ}, which includes contributions from multi-particle states, recall that for bosonic systems with integer-spaced levels we can write
\begin{align}
-\log Z^{(0)}(\beta) &= \sum_{n=1}^{\infty} d_n \log(1-x^n)= \sum_{n=1}^{\infty}\sum_{k=1}^{\infty} \frac{d_n}{k} x^{kn} \nonumber\\
&= \sum_{k=1}^{\infty}  \frac{Z_{\rm SP}(x^k) }{k}
\end{align}
where $Z_{\rm SP}(\beta)$ is the single-particle partition function, with a similar final expression for a fermionic system.  $Z^{(0)}(\beta)$ is only a part of the expression \eqref{eq:FiniteVolumeZ} for $Z(\beta)$, since it leaves out the Casimir vacuum energy.  Hence unless the Casimir energy happens to be zero, $Z^{(0)}(\beta)$ will not enjoy $T$-reflection symmetry.  Indeed, for most QFTs, $Z^{(0)}(\beta)$ is not $T$-reflection symmetric, and the Casimir energy must be included in $Z(\beta) $ to satisfy $T$-reflection, as can be checked for a free scalar field theory on $S^3_R \times S^1_{\beta}$. 

Nevertheless, consider the $N=\infty$ confined-phase gauge theory partition function \emph{without} the vacuum energy contribution\cite{Aharony:2003sx}:
\begin{align}
Z_G(\beta) &:= \exp \left[-\sum_{k=1}^{\infty}  \frac{Z_{\rm ST}(x^k)}{k} \right]  \\
&=\prod_{n=1}^{\infty}\frac{1}{1-z_V(x^{k}) - n_S z_S(x^{k})  +(-1)^k n_F z_F(x^{k})} \nonumber
\end{align}
Since $z_S(1/x) = -z_S(x)$, $z_F(1/x) = -z_F(x)$, and $1-z_V(1/x) = -[1-z_V(x)]$, we see that 
\begin{align}
Z_G(\beta) = e^{i\pi/2} Z_G(-\beta)
\label{eq:GaugeTheoryTReflection}
\end{align}  
with the prefactor obtained from a zeta-function regularization of $(-1)^{\sum_{n=1}^{\infty} 1}$.  So $Z_{G}(\beta)$ enjoys $T$-reflection symmetry.  This is consistent with the general argument for $T$-reflection symmetry after \eqref{eq:FiniteVolumeZ} only if the renormalized Casimir vacuum energy of the $N=\infty$ theory \emph{vanishes}.  

{\bf Vacuum energy}---To check the $T$-reflection prediction we calculate the Casimir vacuum energy $C$
 \begin{align}
C  = \frac{1}{2} \sum_{n=1}^\infty d_n \omega_n 
 \end{align}
with $R\,\omega_n = n/2$ and $d_n $ are drawn from \eqref{eq:ZST}.  The sum is divergent, and must be regularized and renormalized to find the physical value of $C$.  In many QFTs the simplest way to do this is to observe that $C$ is encoded in the behavior of the physical single particle partition function\footnote{A related but different prescription gives the $a$-anomaly coefficient\cite{Herzog:2013ed,*Giombi:2014yra}.}, which for us is $Z_{\rm ST}(y)$, through
\begin{align}
C[y] &= \left[ {1 \over 4R} y {d \over dy} Z_{\rm ST}(y^2)\right] ={1\over 2} \sum_{n=1}^{\infty} d_n \omega_n y^n.
\label{eq:Cy}
\end{align}
Then $C$ would normally be given by the finite part of $C[y \to 1]$ in the simple class of theories we work with, which have no microscopic mass terms.  This amounts to defining $C$ via an especially natural analytic continuation, and also resembles a heat-kernel regularization, since it effectively introduces the damping factor $e^{-\omega_n/\mu}$, with $\mu = 1/\beta$ playing the role of the UV cutoff.  If we were dealing with a system where $d_n \to q \, n^{p}$ once $n\gg 1$ for some fixed $p, q \in \mathbb{R}^{+}$, then $C[y]$ would be well-defined for any $y \in [0,1)$, and we would expect to find
\begin{align}
C[y\to 1] = R^{3}\mu^{4} + R\mu^2 + C +\mathcal{O}(\mu^1)
\label{eq:expected_divergence}
\end{align}
and the leading power of $\mu$ is tied to the spacetime dimension $d=4$.   Here, the thermodynamic degeneracy factors $d_n$ from \eqref{eq:ZST} are associated with confining large $N$ gauge theories, and it is known that $d_n$ grows \emph{exponentially} with $n$, $d_n \sim p \, n^q \,h^{n}, n\gg 1$ with $p, q, h \in \mathbb{R}^{+}$ and $h>1$.  This is the famous Hagedorn scaling of the density of states.  Consequently, if we keep $\mu \in \mathbb{R}^{+}$,  $Z_{\rm ST}(\mu)$ is only well-defined for $\mu < \mu_H$.  Physically, if the temperature is increased past $T_H$ there is a Hagedorn instability, and a consequent phase transition to a deconfined phase.  So at first glance it is not clear how to use \eqref{eq:Cy} to compute $C$ for confining large $N$ theories.

To circumnavigate this roadblock, note that we do not have to take the $y \to 1$ limit of $Z_{\rm ST}$ along the real axis.  We can approach $y=1$ along any smooth path in the complex plane which does not go through any singularities.  The singularities of $Z_{\rm ST}[y]$ are set by the roots of\begin{align}
p[y] &= 1-z_V(y^{2}) \pm n_F z_F(y^{2}) - n_S z_S(y^2) 
\end{align}
If $p[y]$ has a root $y_H \in [0,1]$, then the logarithms in \eqref{eq:ZST} (which depend on $p[y^k]$) become singular at $y=y_H, y_H^{1/2},y_H^{1/3} \ldots$, and \eqref{eq:ZST} ceases to be well-defined for $y\ge y_H$.  Such roots are present for any integer $n_F, n_S \ge 0$, which is the origin of the Hagedorn instability.    Figure~\ref{fig:YMPlots} shows the location of the singularities of the Yang-Mills (left) and $N_f=1, N_s=2$ (right) single-trace partition functions as red dots, with the blue curve illustrating an example of one of the many approach trajectories to $y=1$ along which there are no singularities.

Armed with this observation, we can evaluate $C$ numerically or analytically.  Let us begin with the analytical approach. The first step to isolate the part that diverges as $y\rightarrow1$ from the rest in $y\, dZ_{ST}/dy$ in \eqref{eq:Cy}; 
\begin{align}
&y{ \partial \over \partial y}  \log \left[1-z_V(y^{2m}) +n_F(-1)^m z_F(y^{2m})- n_S z_S(y^{2m}))\right] \nonumber \\
&=\frac{2 m y^{2 m} \left(3 y^{4 m}-2 (n_S+3) y^{2 m}+6 n_F (-y)^m-n_S-3 \right)}{ y^{6 m}- (3 + n_S) y^{4 m}+ 4 n_F (-y)^{3m}-(3 + n_S )y^{2 m} +1 }\nonumber\\
&\quad+\frac{6 m y^{2 m}}{1-y^{2 m}}=3m+ \frac{6 m y^{2 m}}{1-y^{2 m}}\,,
\label{eq:separation}
\end{align}
where in the last step we substituted $y=1$ in the finite term. This substitution should be understood as a limit in the complex plane that avoids any singularities along its path, as described above. By using Eqs.~\eqref{eq:ZST}, \eqref{eq:Cy}, and \eqref{eq:separation} we obtain the formally divergent expression 
\begin{align}
C=-{3\over 4R}\left(\sum_{m=1}^\infty \varphi(m)  +2\, \lim_{\beta\to0} \sum_{m=1}^\infty {\varphi(m)\, y^{2m}\over1-y^{2m}} \right)\,.
\end{align}
After regulating the first term using a spectral zeta function via the identity  $\sum_{m=1}^\infty \varphi(m) m^{-s}=\zeta(s-1)/\zeta(s)$, and regulating the second using the Lambert series, $\sum_{m=1}^\infty\varphi(m) q^m/(1-q^m)=q/(1-q)^2$, we obtain
\begin{align}
C=-{1\over 4R}\left({3\zeta(-1)\over\zeta(0)} +\frac{6 R^2}{\beta ^2}-\frac{1}{2}\right)=-{3 R\over 2\beta^2}\,
\label{eq:CValue}
\end{align}
up to $\mathcal O(\beta^2)$.   The divergent contribution is cancelled by a $\int d^{4} x \sqrt{g}\, \mathcal{R}$ counter-term, and the lack of a finite term in \eqref{eq:CValue} means that the renormalized $C$ is zero.  A similar calculation gives $\gamma=-3\pi/2$.  At first glance,  splitting terms in \eqref{eq:Cy} and regularizing them individually might seem worrisome, but since we have used a spectral zeta function and the cutoff functions depend only on the spectrum throughout, these manipulations are justified.  

To compute $C$ numerically we examine the $\beta \to 0$ limit of $Z_{\rm ST}[e^{-\beta e^{-i \alpha} }]$ with a cutoff $k_{\rm max}$ on the $k$ sum.  Increasing $k_{\max}$ allows probing $Z_{\rm ST}$ at lower $\beta$.  The leading divergence in $Z_{\rm ST}$ as $\beta \to 0$ turns out to scale as $1/\beta$, rather than $1/\beta^{3}$ as one might have expected from \eqref{eq:expected_divergence}\footnote{A similar result was discussed in \cite{DiPietro:2014bca}.    
}.   The coefficient of $\beta$ in a small-$\beta$ expansion of $Z_{\rm ST}$ is $C$.  Extracting this coefficient using least-squares fits in Yang-Mills theory gives e.g. $C R = (-0.95 +2.22 i)\times 10^{-4}$ with $k_{\rm max} = 2.5 \times 10^5, \alpha = \pi/6$ from sampling $Z_{\rm ST}$ in the range $\beta \in [5 \times 10^{-5}, 6.5\times 10^{-2}]$. $C$ decreases as $k_{\max}$ is increased, which is consistent with the analytic result $C=0$.  We have also checked that the analytic results for $C$ and $\gamma$ (up to branch choice ambiguities) are reproduced numerically for theories with $N_f =0, N_s \ge 0$.  We have not succeeded in getting stable numerical results for $C$ once $N_f \ge1$, so for this subclass of theories our conclusions rely on our two analytic arguments.

{\bf Conclusions}---The confining-phase Casimir vacuum energy in non-supersymmetric large $N$ gauge theories with adjoint matter turns out to be zero.   This result cannot be attributed to cancellations between bosons and fermions, since it holds even in Yang-Mills theory, which has a purely bosonic spectrum.  Since we find a zero vacuum energy in a variety of examples, it is unlikely to be an accident.  It appears that there is a mechanism other than SUSY that can make vacuum energies vanish, at least in a class of $N = \infty$ gauge theories, and consequently also in their string duals. 

Obviously the most pressing task suggested by our results is to understand them in terms of some symmetry principle.  This may involve some novel  emergent large $N$ symmetry of confined phases of gauge theories, or some previously unrecognized $N = \infty$ consequence of an already known symmetry, such as center symmetry.  It will be valuable to gather further clues by generalizing the analysis, and to explicitly compute $1/N$ corrections to the vacuum energy.  Depending on how broadly the results generalize, it is possible that they may find phenomenological applications.   It is important to see whether the vacuum energy continues to vanish if additional scales are introduced into the problem, for instance by working with a squashed $S^3$, and to understand the consequences of including contributions from other matter field representations.  Finally, we note that there may be some relations between our results and the recent observation that the  $S^3\times S^1$ Casimir energy vanishes in non-interacting conformal higher-spin theories\cite{Beccaria:2014jxa}.

{\it Acknowledgements:}  We thank O. Aharony, J. L. Evans, C. P. Herzog, K. Jensen, I. Klebanov, M. \"Unsal, and H. Verlinde for discussions.  This work is supported in part by the U. S. Department of Energy under the grants DE-FG-88ER40388 (G.~B.) and DE-SC0011842 (A.~C.). 

\bibliographystyle{apsrev4-1}
\bibliography{Treflection.bib} 
\end{document}